\documentclass{article}
\usepackage[T1]{fontenc} 
\usepackage[utf8]{inputenc} 
\usepackage{ismir,amsmath,cite,url}
\usepackage{graphicx}
\usepackage{color}

\usepackage{lineno}

\title{DadaGP: A Dataset of Tokenized GuitarPro Songs for Sequence Models}

\multauthor
{Pedro Sarmento$^1$ \hspace{1cm} Adarsh Kumar$^{2,3}$ \hspace{1cm} CJ Carr$^4$} { \bfseries{Zack Zukowski$^4$ \hspace{1cm} Mathieu Barthet$^1$ \hspace{1cm} Yi-Hsuan Yang$^{2,5}$}\\
  $^1$ Queen Mary University of London , UK \hspace{.5cm}
$^2$ Academia Sinica, Taiwan\\
$^3$ Indian Institute of Technology Kharagpur, India \hspace{.5cm}
$^4$ Dadabots \hspace{.5cm} $^5$ Taiwan AI Labs \\
{\tt\small\{p.p.sarmento, m.barthet\}@qmul.ac.uk, emperorcj@gmail.com, yhyang@ailabs.tw}
}

\def\authorname{P. Sarmento, A.Kumar, CJ Carr, Z. Zukowski, M. Barthet and Y. Yang}

\PassOptionsToPackage{hyphens}{url}\usepackage[bookmarks=false,pdfauthor={\authorname},pdfsubject={\papersubject},hidelinks]{hyperref}

\begin{document}

\maketitle
\begin{abstract}
Originating in the Renaissance and burgeoning in the digital era, tablatures are a commonly used music notation system which provides explicit representations of instrument fingerings rather than pitches. GuitarPro has established itself as a widely used tablature format and software enabling musicians to edit and share songs for musical practice, learning, and composition. In this work, we present DadaGP, a new symbolic music dataset comprising 26,181 song scores in the GuitarPro format covering 739 musical genres, along with an accompanying tokenized format well-suited for generative sequence models such as the Transformer. The tokenized format is inspired by event-based MIDI encodings, often used in symbolic music generation models. The dataset is released with an encoder/decoder which converts GuitarPro files to tokens and back. We present results of a use case in which DadaGP is used to train a Transformer-based model to generate new songs in GuitarPro format. We discuss other relevant use cases for the dataset (guitar-bass transcription, music style transfer and artist/genre classification) as well as ethical implications. DadaGP opens up the possibility to train GuitarPro score generators, fine-tune models on custom data, create new styles of music, AI-powered songwriting apps, and human-AI improvisation.
\end{abstract}




\section{Introduction}\label{sec:i}



\begin{figure}\label{fig:tab}
 \hspace{1mm}
 \includegraphics[width=0.9\columnwidth]{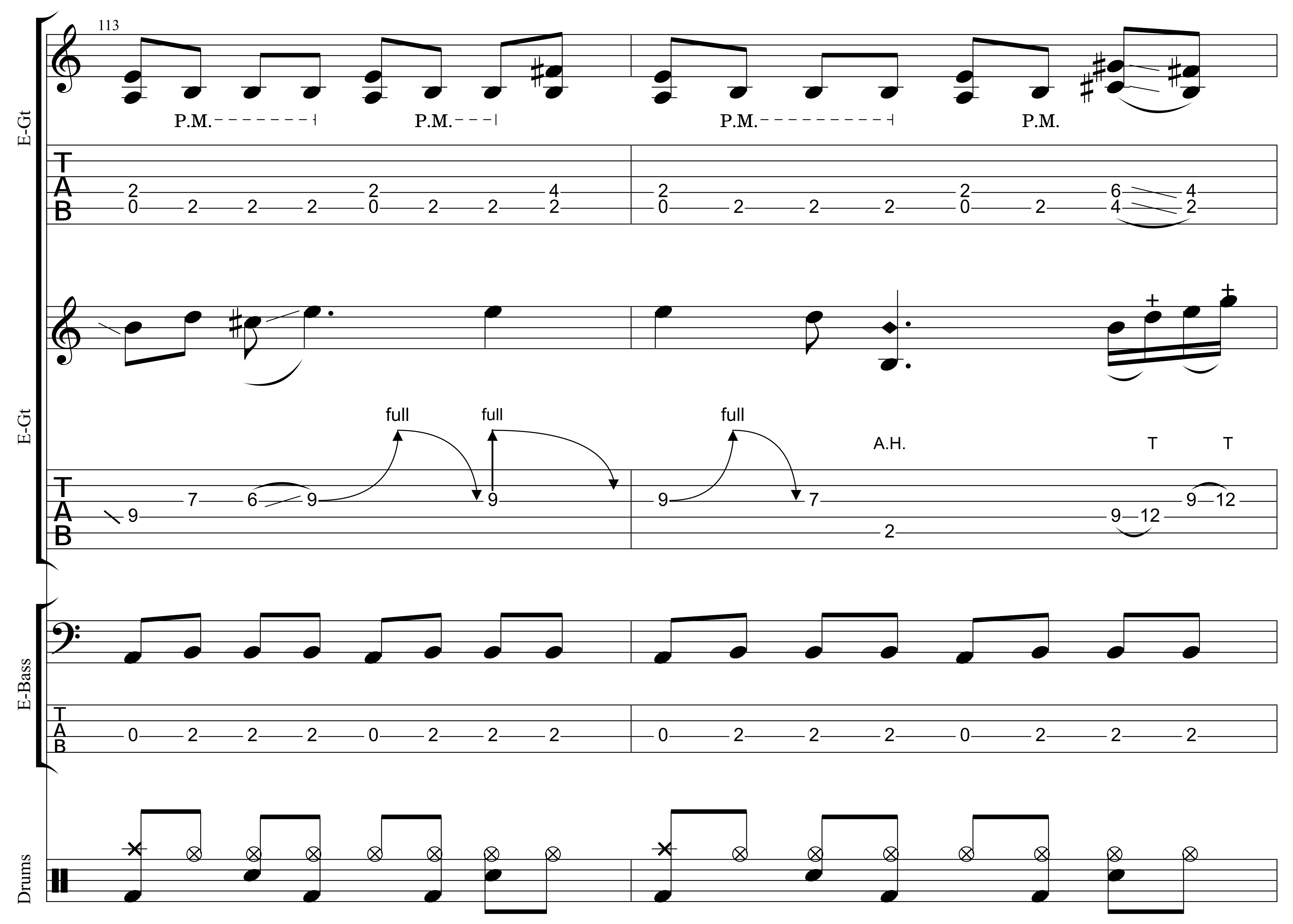}
 \caption{An excerpt from a GuitarPro song notation using tablatures and score for two guitars, bass and drums.}
 \label{fig:example}
\end{figure}

Historically, tablatures' proliferation is closely linked to the lute repertoire, compositions that roughly span from the 16th century onwards, and are still available today \cite{DeValk2019}. In opposition to standard notational practices (usually referred to as staff notation), in a tablature system for string instruments each staff line on the score represents a string of the instrument, substituting a representation of pitch by a given location on said instrument (i.e. a fingering) \cite{Magnusson2019}. Tablatures are a \textit{prescriptive} type of notation, where the emphasis is on the action (symbol-to-action), contrary to \textit{descriptive} forms of notation, which establishes a symbol-to-pitch relationship. This characteristic makes tablatures an intuitive and inclusive device for music reading and learning, which can explain their large prevalence for music score sharing over the Internet \cite{Macrae2011,barthet2011}. Often represented as non-standardised text files that require no specific software to read or write, tablatures' online dissemination has surpassed more sophisticated music notation formats, such as Music XML or MIDI \cite{Macrae2011}. However, tablature representations that rely solely on text have limitations from a user perspective. For example, it is common that rhythm indications are discarded, preventing a comprehensive transcription of the music and automatic playback. Tablature edition software (e.g. GuitarPro\footnote{\url{https://www.guitar-pro.com/}}, PowerTab\footnote{\url{http://www.power-tab.net/guitar.php}}, TuxGuitar\footnote{\url{https://sourceforge.net/projects/tuxguitar/}}) can be regarded as a solution for this problem, keeping the \textit{prescriptive} approach, and supporting rhythm notations and playback.  By supporting the annotation of multiple instruments, as observable in Figure \ref{fig:tab}, these tools account for an interactive music experience, either for songwriting or music learning purposes.

The release of this dataset intends to leverage the GuitarPro format used by the before-mentioned software to support guitar and bands/ensembles' related research within the MIR community, focusing specifically on the task of symbolic music generation. 
The contributions of this paper are: (1) a  dataset of over 25,000 songs in GuitarPro and token format, together with statistics on its features and metadata, (2) an algorithm and Python software to convert between any GuitarPro file and a dedicated token format suitable for sequence models\footnote{Available at: \url{https://github.com/dada-bots/dadaGP}}, (3) results from its main use case, the task of symbolic music generation, and (4) a discussion about further applications for DadaGP and its ethical implications.

In this paper, we first present some relevant background concerning previously released music datasets in symbolic format. In Section \ref{sec:wganm}, we discuss advantages of tab-based datasets for MIR research. We then describe, in Section \ref{sec:dataset}, the details of the DadaGP dataset, its encoder/decoder support tool, the features it encompasses and the ones it lacks. Within Section \ref{sec:ucamg} we present a use case of symbolic music generation using our proposed dataset, supported by previous approaches concerning databases of symbolic music. Section \ref{sec:pa} proposes additional applications for the dataset. Finally, in Section \ref{sec:d} we explain the steps needed in order to acquire the dataset, further pointing out some ethical considerations in Section \ref{sec:ec}.

\section{Background}\label{sec:b}

Since its release in 1983, the MIDI (Music Instrument Digital Inferfaces) standard has remained highly ubiquitous. Unsurprisingly, MIDI has been the most recurrent option in terms of musical notation formats, concerning datasets released within the MIR community, either targeting music generation purposes, that lately have boomed by leveraging deep learning approaches, or aiming for musical analysis, musicology or purely information retrieval ends. A comprehensive overview of previously released datasets in symbolic format is presented in \cite{Dong2020}. The authors present MusPy, a toolkit for symbolic music generation, that natively supports a total of eleven datasets. Considering cumulative song duration, the top five datasets are the Lakh MIDI dataset \cite{Raffel2016}, the MAESTRO dataset \cite{Hawthorne2019}, the Wikifonia Lead Sheet dataset\footnote{No longer available.}, the Essen Folk Song database \cite{Essen}, and the NES Music database \cite{Donahue2018}. With respect to music notation formats, these datasets employ MIDI, MusicXML and ABC. Recently, 
the GiantMIDI-Piano dataset \cite{Kong2020}, comprising 10,854 unique piano solo pieces,
the POP909 dataset \cite{Wang2020} and the Ailabs.tw Pop1K7 dataset \cite{Hsiao2021}, containing respectively piano arrangements of 909 and 1,748 popular songs,
were also released,
all relying on MIDI format. This standardisation around MIDI is useful for there are several Python libraries to work with this format, such as  music21 \cite{Cuthbert2010}, mido \cite{mido}, pretty\_midi \cite{Raffel2014}, and jSymbolic \cite{Mckay2006}. 

Regarding guitar-oriented research, previous dataset releases have not particularly targeted automatic music generation  goals, instead focusing on guitar transcription or playing technique detection. The GuitarSet consists of 360 excerpts of acoustic guitar along with annotations for string and fret positions, chords and beats \cite{Xi2018}. Furthermore, the Guitar Playing Techniques dataset \cite{Su2014} contains 6,580 clips of notes together with playing technique annotations. Likewise, the IDMT-SMT-Guitar dataset \cite{Kehling2014} also comprises short excerpts that include annotations of single notes, playing techniques, note clusters, and chords. Lately, Chen et al. compiled a dataset of 333 tablatures of fingerstyle guitar, created specifically for the purpose of music generation \cite{Chen2020}.

To the authors best knowledge, there exists no multi-instrument dataset that is able to combine the ease of use of symbolic formats whilst providing guitar (and bass) playing technique information. Such expressive information is lacking in other formats, and GuitarPro appears as a viable resource for music analysis and generation.

\section{Motivations: Why GuitarPro?}\label{sec:wganm}
GuitarPro is both a software and a file format, widely used by guitar and bass players, but also by bands. It is mostly utilized for tasks such as music learning and practicing, where musicians simply read or play along a given song, and for music notation, in which composers/bands use the software to either support the songwriting process, or simply as a means for ease of distribution once compositions are done. As an example of the software's widespread dissemination, the tablature site Ultimate Guitar\footnote{\url{https://www.ultimate-guitar.com/}} hosts a catalogue of over 200,000 user-submitted GuitarPro files, containing notations of commercial music, mostly from the genres of rock and metal. 
One of the main motivations for the creation of DadaGP is to engage the MIR community into research that leverages the expressive information, instrumental parts and song diversity in formats such as GuitarPro. Although GuitarPro is a paid software, free alternatives such as TuxGuitar are capable of editing/exporting into GuitarPro format. Moreover, GuitarPro files can be easily imported into MuseScore\footnote{\url{https://musescore.com/}}, a free software notoriously known for music notation, which also possesses tablature features. However, using MuseScore might present some occasional incompatibilities, specifically those regarding the selection of instruments (e.g. drums are often imported as piano, and the corresponding MIDI instruments need to be manually switched). Another important motivation for the release of this dataset is that it is possible to make conversions between GuitarPro and MIDI files. This can be done inside any of the aforementioned software, by simply exporting into MIDI, or by scripting. Thus, by converting the dataset's GuitarPro files into MIDI, MIDI-based music feature extraction functions available (e.g. Python libraries referenced in Section \ref{sec:b}) can be applied. 
Finally, we believe that our dataset is able to provide researchers with the information present in standard MIDI datasets, while including at the same time prescriptive information useful for guitar-oriented research.

\section{DadaGP Dataset}\label{sec:dataset}


Leveraging the proliferation of music transcriptions available online as GuitarPro files, we compiled DadaGP, a dataset containing 26,181 songs. We also devised an enconding/decoding tool to convert GuitarPro files into tokens, which is described in Section \ref{ssec:edt}. 
In total, it contains 116M tokens, which is about the size of WikiText-103 \cite{DBLP:journals/corr/MerityXBS16}. In terms of duration, the dataset amounts to over than 1,200 hours (average song length of 2:45 minutes).


\subsection{Encoding/Decoding Tool}\label{ssec:edt}

\subsubsection{Feature Extraction with PyGuitarPro}
PyGuitarPro \cite{PyguitarPro} is a Python library which reads, writes and manipulates GuitarPro files\footnote{Currently, it supports GP3, GP4 and GP5 files.}. Our encoding/decoding tool explores its feature extraction functions, in order to  convert much of the information into a tokenized text format. With PyGuitarPro it is possible to acquire information regarding music-theoretic features (e.g. pitch, rhythm, measure, instrument) and playing technique information.

\subsubsection{Tokenization}

\begin{figure}[ht]
 \hspace{0mm}
 \includegraphics[width=0.9\columnwidth]{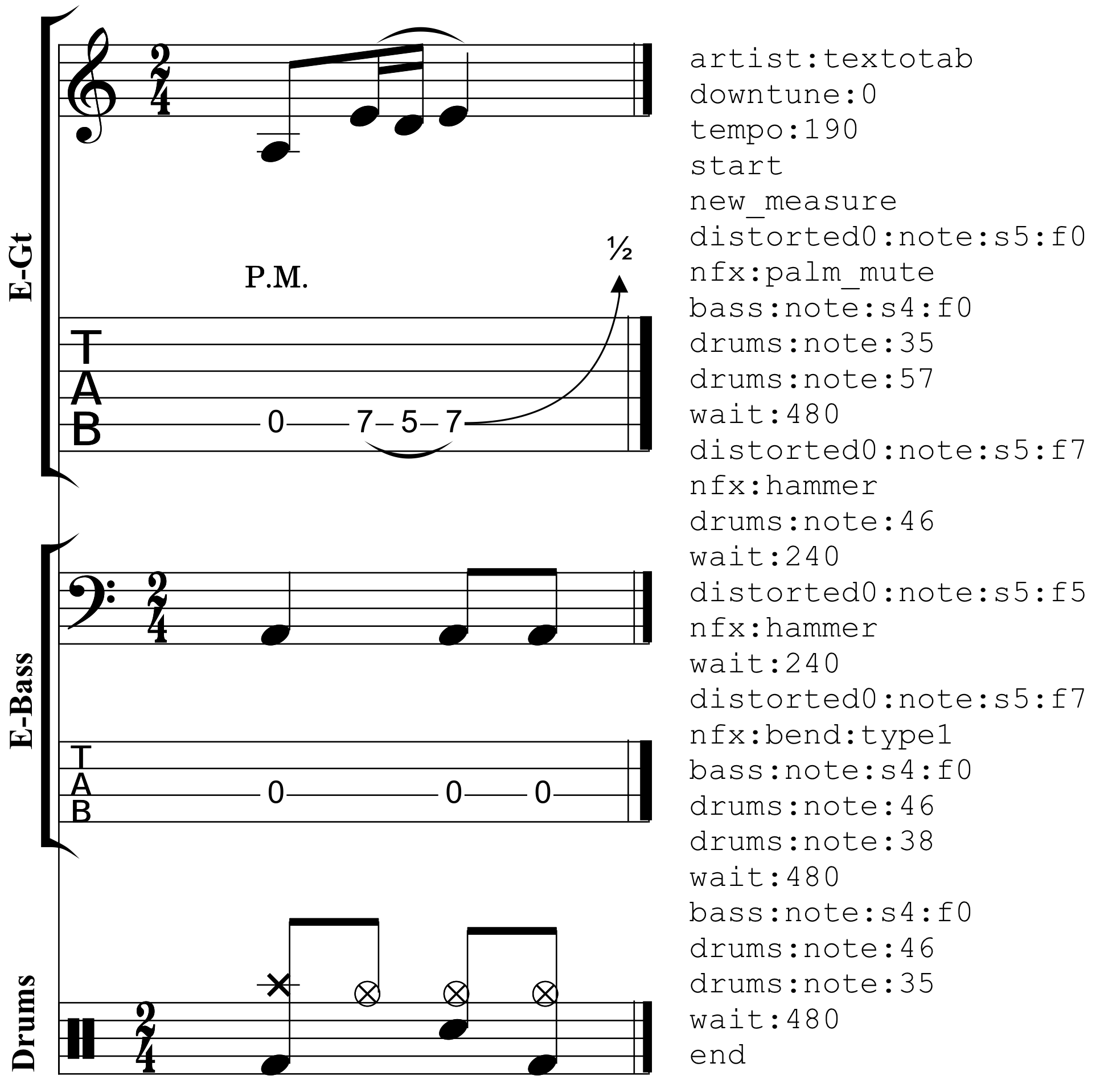}
 \caption{A measure with a distorted guitar, bass and drums in GuitarPro's graphical user interface (left), and its conversion into token format (right).}
 \label{fig:textotab}
\end{figure}

The token format takes inspiration from event-based MIDI encodings used in previous music generation works, such as MuseNet\cite{christine_2019}, REMI \cite{Huang2020} and CP \cite{Hsiao2021}. The tool consists of a Python script that utilizes PyGuitarPro to process GuitarPro files into/from token format. Syntactically, every song begins with \verb|artist|, \verb|downtune|, \verb|tempo| and \verb|start| tokens. A depiction of the conversion process can be seen in Figure \ref{fig:textotab}. Notes from pitched instruments are represented by a combination of tokens in the format of \verb|instrument:note:string:fret| and rests by \verb|instrument:note:rest|. For the drumset, the representation is done by \verb|drums:note:type|, leveraging GuitarPro 5 percussion MIDI maps (e.g. \verb|drums:note:36| for a kick drum, \verb|drums:note:40| for a snare). Every note or rest is separated in time by \verb|wait| tokens. This is sufficient for the decoder to figure out note durations. There is no need to use note-off tokens, because new notes silence old notes, unless a \textit{ghost note} or \textit{let ring} effect is used. Every new measure, note effect, beat effect, and tempo change is registered as a token. Effect tokens are applied to the preceding note token. A histogram containing the most common tokens in DadaGP is available in Figure \ref{fig:plots}(g).

Furthermore, the DadaGP token format is resilient to syntax errors, such that random token sequences will still produce decodable music. We believe this is helpful when creatively pushing generators to make out-of-distribution sequences using high temperatures, early epochs, extreme latent dimension values, interpolated style conditioning, and other experimental practices.

\subsection{Repertoire}\label{subsec:r}
\begin{figure}[ht]
 \hspace{-1mm}
 \includegraphics[width=1\columnwidth]{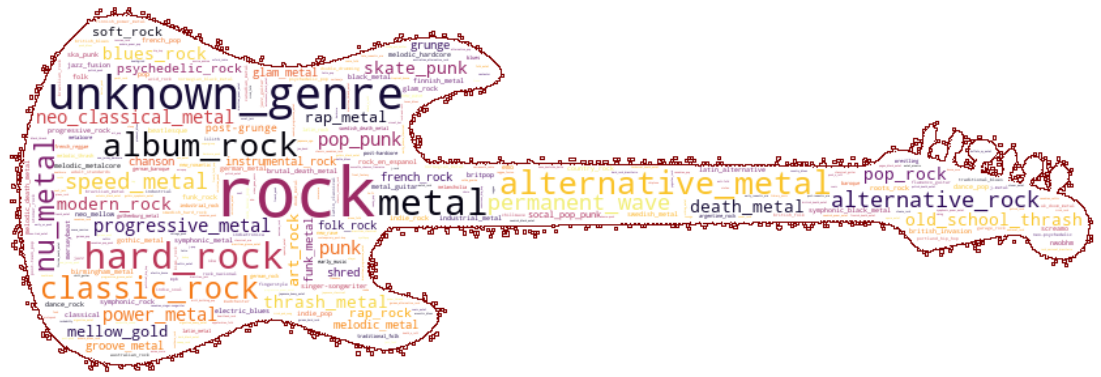}
 \caption{Word cloud representation of the musical genres in DadaGP. Tag size increases with amount of songs.}
 \label{fig:genres}
\end{figure}

\begin{figure*}[!ht]
 \includegraphics[width=1\textwidth]{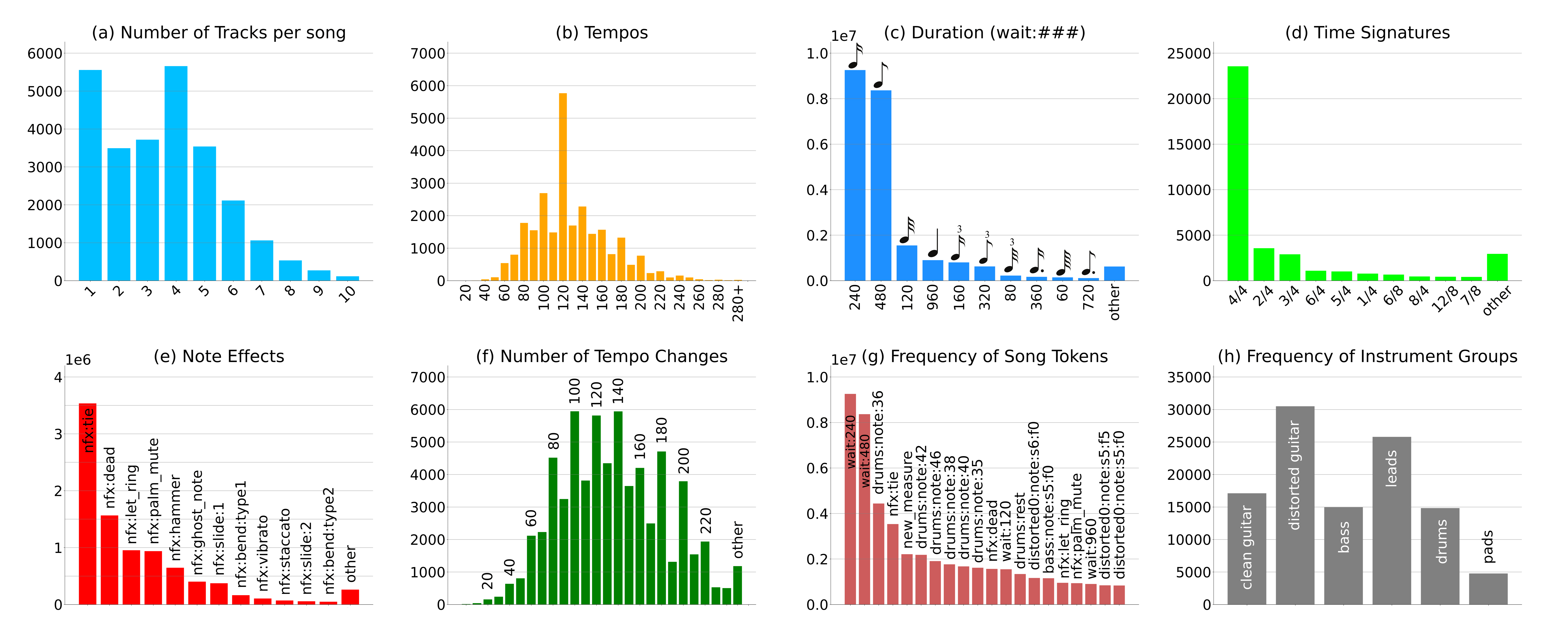}
 \caption{Statistical information about the DadaGP dataset. Histograms of tracks per song (a), initial tempos (b), most common note durations in token and staff notation format (c), time signatures (d),  note effects (e), amount of tempo changes (f), most frequent tokens (g) and instruments (h).}
 \label{fig:plots}
\end{figure*}

Each song is labelled with artist and genre information, although genre tags are absent within original GuitarPro files. To this end, we compiled a genre list, with information acquired from the Spotify Web API\footnote{Available at: \sloppy\url{https://developer.spotify.com/documentation/web-api/}}, querying by artist and song title, resulting in genre metadata for each composition. It is worth mentioning that a given song can have more than one genre attached to it. Information about the most prevalent genres within DadaGP can be seen in Figure \ref{fig:genres}. While its emphasis is on genres and sub-genres from rock and metal, its corpus is diverse, also including stylistically distinct genres such as jazz, classical, pop and EDM. From Figure \ref{fig:plots}(a) we observe that most of the songs in DadaGP contain four instrumental parts, usually two guitars, a bass and drums.
 
\subsection{Instruments}
Regarding instrumentation, for DadaGP a maximum of nine instruments were chosen: three distorted or overdriven guitars, two clean or acoustic guitars, one bass, one drumset, one lead (for instruments with sharp attacks, e.g. piano), and one pad (for instruments used more ambiently, like a choir or a string ensemble). Multiple drum tracks are combined into one. Rare instruments are combined into the lead and pad tracks. In Figure \ref{fig:plots}(h) we can notice a predominance of distorted guitars in the dataset. Intuitively this is justified by the presence of two distorted guitars (often one rhythmic and one lead) on most of the songs in DadaGP, due to the predominance of the rock/metal genre.   
Concerning guitar and bass, 7 string guitars are supported, as are 5 and 6 string basses. Downtuning is supported only if all instruments downtune the same amount, and common tunings such as \textit{Drop D}\footnote{A tuning in which only the lowest string is downtuned by one whole step, usually from E to D.} and \textit{Drop AD} are also included. Rare tunings were dropped from the dataset as the encoder does not support them. 

Guitar playing technique notations are represented by note effect tokens (\verb|nfx|), although this family of tokens also holds information about other instruments (e.g. \verb|nfx:tie|, which acts as a link between two adjacent notes). On Figure \ref{fig:plots}(e) we present a histogram of the most frequent occurrences of these in our dataset, namely \textit{palm mute} (a technique often used with distortion guitars where the guitar player dampens the strings with the right hand palm), bends and vibratos, slides, hammer-ons and pull-offs (both under \verb|nfx:hammer|).

\subsection{Meter}

As clarified before, each note/rest event is followed by a \verb|wait| token which specifies the number of ticks between it and the succeeding event. In DadaGP, tick resolution uniformly corresponds to 960 ticks per quarter note. For a tempo of 100 bpm, a tick corresponds to $60 / (100*960) = 0.000625$ seconds. Referring to the excerpt in Figure \ref{fig:textotab}, eighth note events are separated by \verb|wait:480| tokens, and sixteenth note ones by \verb|wait:240|. A histogram with the most common durations in DadaGP is presented in Figure \ref{fig:plots}(c), in both token and standard staff notation formats, to ease visualization. 

Usually, in a GuitarPro file a default tempo is specified for the entire song, although it supports the inclusion of tempo changes throughout the piece. This is addressed by our encoder/decoder with the tokens \verb|tempo| and \verb|bfx:tempo_change| respectively, which affects note/rest duration. In Figure \ref{fig:plots}(b) and Figure \ref{fig:plots}(f) are presented plots corresponding to the most frequent tempos and tempo changes.

The encoder/decoder also supports the representation of measure repetitions with the \verb|measure:repeat| token. Although time signatures are not tokenized, they are inferred by summing the \verb|wait| tokens between the occurrences of \verb|new_measure|. However, this method is insufficient to distinguish between $3/4$ and $6/8$ measures, for example. To circumvent this, for the plot presented in Figure \ref{fig:plots}(d) we leveraged PyGuitarPro functions to extract accurate information about the most prevalent time signatures for each measure in our dataset.

\subsection{What is Missing?}

Information regarding key signature is not provided as part of the dataset. Although key signature can be represented in GuitarPro format, it is rarely present within files. Similarly to the results presented in \cite{Raffel2016} for the Lakh MIDI dataset, 93.7\% of the songs in DadaGP were automatically assigned the key of C Major, rendering these statistics inaccurate.

GuitarPro does not include note velocity information as in MIDI. However, in GuitarPro loudness between notes and musical phrases is notated using traditional dynamic instructions (e.g. \textit{forte}, \textit{pianissimo}, \textit{mezzo-forte}). In its token format, DadaGP does not yet support this, but there is a possibility of accentuating notes at two levels with \verb|nfx:accentuated_note| and \verb|nfx:heavy_accentuated_note|. 

Concerning vocals, a common practice with GuitarPro files is to select MIDI wind instruments to notate singing melodies. Currently, our dataset is not well-suited to handle vocals, for these get converted into the \verb|leads| instrument, which may also contain information about other instruments, such as the piano. Lyrics are also possible to include in GuitarPro, but that feature is currently not supported by our encoder/decoder tool.

\section{Use Case: Symbolic Music Generation}\label{sec:ucamg}

Recently, the field of symbolic music generation has witnessed consistent progress. Considering works that target symbolic music generation with Transformer-based models, MusicTransformer \cite{AnnaHuang2019} is a MIDI generator trained on piano performances with improved long-term coherence over vanilla RNNs due to the use of the Transformer \cite{Vaswani2017}. Similarly, MuseNet \cite{christine_2019} is a generative Sparse Transformer \cite{Child2019} trained on a larger dataset of MIDI including over 600 styles. An API for the model was launched by OpenAI, which powers the songwriting app MuseTree \cite{musetree}. However, the model was not released, so it cannot be fine-tuned on custom data. In \cite{metallica} the author trained a charRNN generator on dozens of GuitarPro songs encoded as a sequence of ASCII characters. It only supported one instrument, and its verbose character-sequence format opened up the possibility for syntax errors. 

We tested the DadaGP dataset for a symbolic music generation use case by using the Pop Music Transformer model \cite{Huang2020}, in which the authors devised a Transformer-XL \cite{Dai} architecture to generate pop piano compositions in MIDI format. The reason for the choice of this architecture is because this work considers metrical structure in the input data, allowing for an increased awareness in terms of beat-measure structure. We chose the Transformer-XL model as it is able to learn dependencies that are 450\% longer than vanilla Transformers, thus well-suited for our task. As per the settings, similarly to the original paper, we used $M=8$ attention heads and $N=12$ self-attention layers. 

As a proof-of-concept, we collected a subset from our dataset, retrieving 6,910 songs labelled as \verb|genre:rock|. We generated a list of all the unique tokens in this subset, creating a vocabulary with 2,104 entries. 

\begin{figure}
 \hspace{0mm}
 \centering
 \includegraphics[width=1\columnwidth]{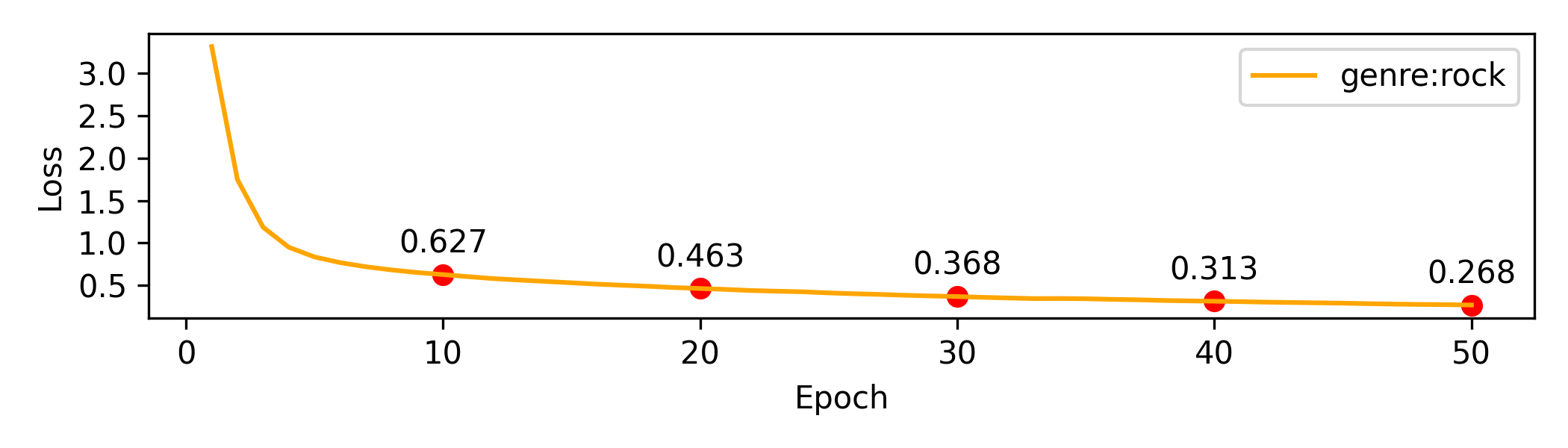}
 \caption{Training loss of the rock subset model, per epoch.}
 \label{fig:loss}
\end{figure}

Training was set to run for 50 epochs. With around 43M parameters, this model took around 10 days to perform this task on a Tesla K80 GPU. We consider this to be impractical in terms of reproducibility, so we intend to release pre-trained models from epochs 40 and 50, for which losses can be seen in Figure \ref{fig:loss}.  

\bigskip
Regarding inference, we conditioned the model by prompting it with an initial list of tokens comprising \verb|artist|, \verb|downtune|, \verb|tempo| and \verb|start|, necessary for the DadaGP decoder. Furthermore, in an attempt to guide the model towards the generation of music comprising specific instruments, we included tokens for a single note of a distorted guitar, bass guitar and drums. Through experimentation, we set on a limit of 1,024 tokens for each generated song, using 1.2 as temperature parameter. Finally, we manually appended an \verb|end| token in order for the decoder to be able to convert it to GuitarPro format, as this is the instruction which tells the decoder when the song finishes.

\begin{figure}
 \hspace{0mm}
 \centering
 \includegraphics[width=0.9\columnwidth]{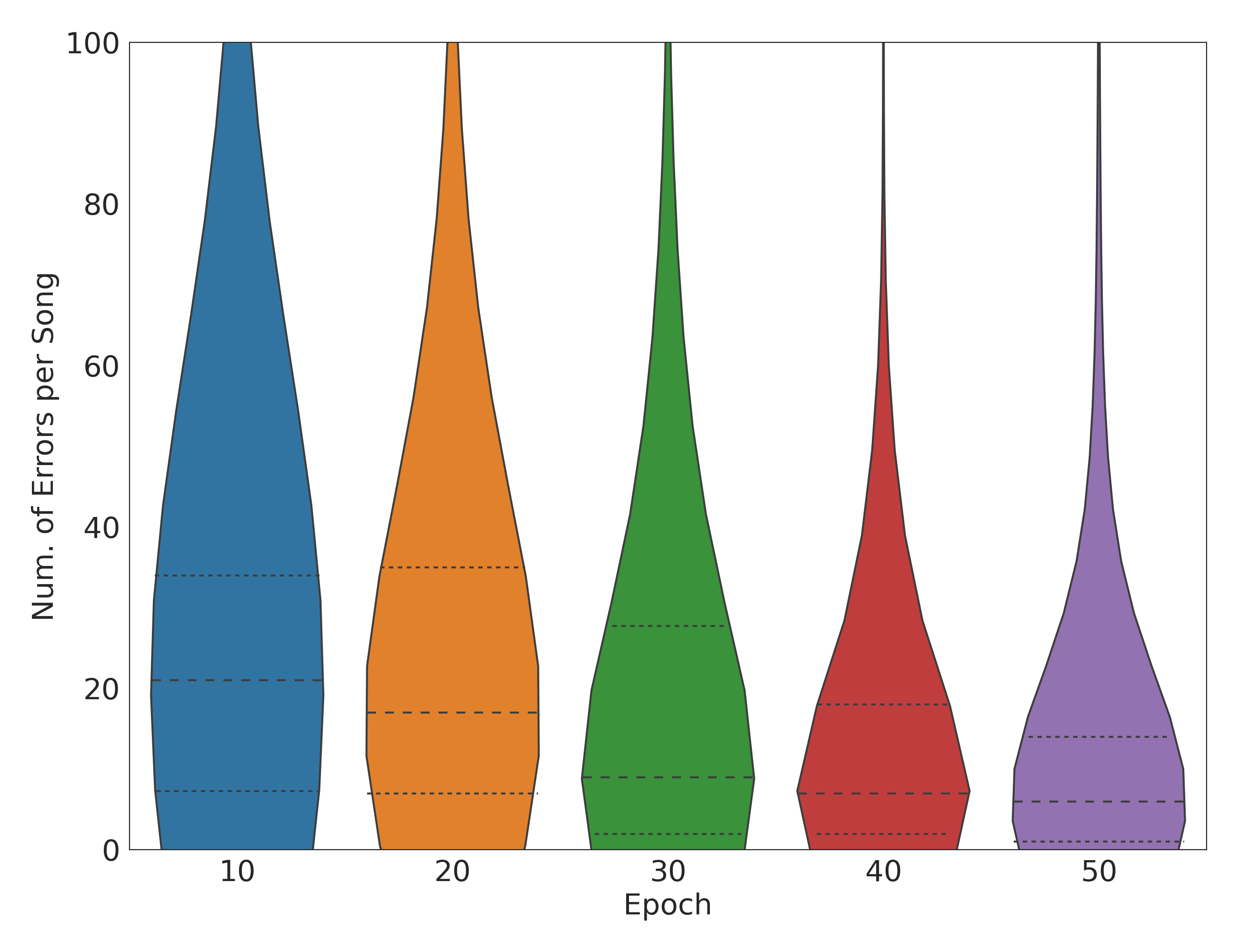}
 \caption{Violin plot of number of errors per song at different epochs.}
 \label{fig:errors}
\end{figure}

As a simple evaluation metric, we focused on the notion of \textit{grammar errors}, namely repetitions of the tokens that should only occur once (\verb|artist:|, \verb|downtune:|, \verb|tempo:|, \verb|start| and \verb|end|), or adjacent repetitions of the same token. Using this, we estimated the number of errors per song, for a corpus of 1,000 generated songs from the model at epochs 10, 20, 30, 40 and 50. As observable in Figure \ref{fig:errors}, not only the median of the number of errors per song is smaller in later epochs, but also the occurrence of outliers is diminished, as expected.

Despite the limitations of the current evaluation, it allowed us to notice a predominance of a specific error, namely the repetition of the token \verb|end|. This is problematic, because the decoder immediately stops the conversion when an \verb|end| token appears, ultimately shortening songs when in GuitarPro format. To counter this effect, we devised a condition that, during inference, would force the model to sample a different token in the event that an \verb|end| token is selected. Results of generated songs without any curation or post-processing have been made available\footnote{Available at: \sloppy\url{https://drive.google.com/drive/folders/1USNH8olG9uy6vodslM3iXInBT725zult?usp=sharing}}.

\section{Prospective Applications}\label{sec:pa}

Although primarily tailored for symbolic music generation, below we describe further applications for DadaGP. 

\subsection{Guitar-Bass Transcription}

The task of guitar-bass transcription from audio recordings is still mostly done manually by musicians, requiring expertise and being both effort and time consuming. In order to automate this, previous research has focused on both solo bass guitar \cite{abeber17taslp, abesser17AES}  and solo guitar \cite{wiggins2019guitar, rodrguez18fma, su19tismir} transcription. As a contribution to solve this problem, we anticipate that DadaGP can be used to create a synthetic dataset for training guitar-bass transcription models, by rendering its corpus from tablatures into audio, using a DAW and appropriate sound fonts. Such a synthetic dataset can be used to pre-train a model, which can then be fine-tuned afterwards using realistic sounds with aligned scores. This argument is supported by the promising results shown by the Slakh dataset, a synthesized version of the Lakh MIDI dataset, on the task of music source separation \cite{manilow2019cutting}. 

\subsection{Music Style Transfer}

Recently, the task of style transfer, the process of changing the style of an image, video, audio clip or musical piece so as to match the style of a given example, has been the subject of much attention. First investigated in applications that target computer vision, music style transfer has recently shown promising results in both the audio \cite{ijcai2019-652} and symbolic domains \cite{brunner,cifka,musemorphose}.
As a prospective application of DadaGP, we envisage that genre information can be leveraged in segregating the dataset across different genres, rendering it suitable for the task of musical genre style transfer, as proposed in \cite{lim} for the specific morphing between Bach chorales and Western folk tunes. Furthermore, besides musical genre, artistic information can also be used towards the task of composer style transfer, once again by filtering DadaGP across distinct artists. 

\subsection{Artist/Genre Classification}

Another task for which artistic and musical genre information present in DadaGP is useful is artist/genre classification. We hypothesize that these features can be used to train classification models, in order to predict composer style and genre related information from the symbolic representation of the songs itself, similarly to what has been implemented in \cite{tsai,sung,kotsifakos}. A thorough survey of the most important approaches regarding music genre classification in the symbolic domain can be consulted in \cite{CORREA2016190}.
Furthermore, there is a symbiosis between this task and the one present in the previous subsection, since the models trained for artist/genre classification can be prospectively used in composer style-based feature extractions, which can be further utilized in tasks like composer style conditioned generation and music style transfer.

\section{Distribution}\label{sec:d}

To ensure reproducibility and facilitate the usage of the dataset, we allow researchers to access DadaGP from a Zenodo repository\footnote{\url{https://zenodo.org}}, on application by request. Here we include the token format versions of the songs, the encoder/decoder Python script in order to convert them into/from GuitarPro format, and the statistical data presented on this paper.

\section{Ethical Considerations}\label{sec:ec}
Training large models has a carbon footprint. Some cloud services are carbon neutral, others are not. This should be considered when training large models on this data. Releasing pre-trained models reduces impact, and we intend to do so with the models present in this paper. 

Many questions regarding production and consumption of music created with AI are still unanswered. For example: Is it wrong to train machine learning models on copyrighted music? Should this be protected by fair use for artists and scientists? What about commercial use? How to acknowledge, reward and remunerate artists whose music has been used to train models? What if an artist does not want to be part of a dataset? Should creators have a monopoly on their style and exclude others from using their style? Or is style communal? Some of these questions were also raised upon the release of Jukebox\cite{Dhariwal2020}, an audio model trained on more than 7,000 artists. However, OpenAI made the case that "\textit{Under current law, training AI systems constitutes fair use (...)}" and  that "\textit{Legal uncertainty on the copyright implications of training AI systems imposes substantial costs on AI developers and so should be authoritatively resolved}"\cite{OpenAI2019}.

\section{Conclusion And Future Work}\label{sec:cafw}
In this paper we presented DadaGP, a dataset of songs in GuitarPro and token formats, together with its encoding/decoding tool. We discussed the features, strengths and weaknesses of the dataset. Moreover, we presented a symbolic music generation use case entailing a novel approach for multi-instrument music generation in tablature format. Finally, we pointed out  additional research applications for DadaGP and discussed some ethical implications.
We intend to improve the DadaGP dataset, namely the possibility of removing \verb|measure:repeat| tokens. During generation, we discovered that these tokens were often hard for the model to interpret, sometimes leading to disproportionate measure repetitions. Also, we plan to include note and phrase dynamics information, and the support for vocal instrumental parts. Regarding music generation, we envision to (1) release a pre-trained model which can be fine-tuned on new music, (2) collaborate with artists that use GuitarPro, (3) explore genre/style transfer, (4) and attempt to play the generated songs in social performances.

\section{Acknowledgments}
This work is supported by the EPSRC UKRI Centre for Doctoral Training in Artificial Intelligence and Music (Grant no. EP/S022694/1). Thanks to Colin Raffel, Brian McFee, and Sviatoslav Abakumov for discussions and advice.


\bibliography{ISMIRtemplate}

%
%
%
%
%

\end{document}